\RequirePackage{fix-cm}
\documentclass[twocolumn,final]{svjour3}          
\smartqed  
\usepackage{graphicx}
\graphicspath{{img/}} 
\DeclareGraphicsExtensions{.pdf,.jpeg,.jpg,.png} 
\usepackage{subfig} 
\usepackage{biograph} 
\usepackage[cmex10]{amsmath}
\usepackage{url}
\usepackage{fancyhdr}
\begin{document}

\title{Optical quorum cycles for efficient communication
	\thanks{Research funded in part by NSF Graduate Research Fellowship Program, IBM Ph.D. Fellowship Program, Symbi GK-12 Fellowship at Iowa State University, and the Jerry R. Junkins Endowment at Iowa State University.  The research reported in this paper is partially supported by the HPC@ISU equipment at Iowa State University, some of which has been purchased through funding provided by NSF under MRI Grant number CNS 1229081 and CRI Grant number 1205413.  Any opinions, findings, and conclusions or recommendations expressed in this material are those of the author(s) and do not necessarily reflect the views of the funding agencies.}
}
\subtitle{\footnotesize The final publication is available at Springer via \url{http://dx.doi.org/10.1007/s11107-015-0561-8}}


\author{Cory J. Kleinheksel         \and
        Arun K. Somani 
}


\institute{Cory J. Kleinheksel \and Arun K. Somani \at
              Iowa State University, Ames, IA 50011 \\
              Tel.: 515-294-0442, Fax: 515-294-9273\\
              \email{\{cklein,arun\}@iastate.edu}           
}

\date{Received: date / Accepted: date}

\maketitle

\begin{abstract}
Many optical networks face heterogeneous communication requests requiring topologies to be efficient and fault tolerant.  For efficiency and distributed control, it is common in distributed systems and algorithms to group nodes into intersecting sets referred to as quorum sets.  We show efficiency and distributed control can also be accomplished in optical network routing by applying the same established quorum set theory.

Cycle-based optical network routing, whether using SONET rings or p-cycles, provides the sufficient reliability in the network.  Light-trails forming a cycle allow broadcasts within a cycle to be used for efficient multicasts.  Cyclic quorum sets also have all pairs of nodes occurring in one or more quorums, so efficient, arbitrary unicast communication can occur between any two nodes.  Efficient broadcasts to all network nodes are possible by a node broadcasting to all quorum cycles to which it belongs ($O(\sqrt{N})$).

In this paper, we propose applying the distributed efficiency of the quorum sets to routing optical cycles based on light-trails.  With this new method of topology construction, unicast and multicast communication requests do not need to be known or even modeled a priori. Additionally, in the presence of network link faults, greater than 99\% average coverage enables the continued operation of nearly all arbitrary unicast and multicast requests in the network.  Finally, to further improve the fault coverage, an augmentation to the ECBRA cycle finding algorithm is proposed.
  
\end{abstract}

\section{Introduction}
\label{intro}
We developed a novel method to enhance fault tolerance capabilities of cycles in a network.  The cycle-based routing algorithm is to be used in the optical networks to achieve routing for both point-to-point and multi-point communication. The actual cycles are created using quorums. Within a cycle, multicasts to all nodes in that cycle is possible. The quorum intersection property and the use of cyclic quorums sets provide all of the unicast capabilities.  Exploiting the same properties, we can achieve efficient broadcasts with $O(\sqrt{N})$ multicasts.  These are significant results for modern and future optical networks facing dynamic heterogeneous traffic requests.

Fiber-optic lines make up the foundation of many networks across the globe.  Some networks stretch hundreds of kilometers, while others are contained within buildings or rooms.  These optical circuits are depended upon for high-speed communications in distributed algorithms, as much as they are needed for the arbitrary point-to-point communications.

Failures within a network are to be expected and can happen as much as every couple days \cite{dlastine2012fault}.  Protecting against these optical circuit faults is critical and there are many different approaches depending on the network needs and individual circumstances.  SONET rings can be used to protect point-to-point and shared paths while enabling failure location.  Using a pre-configured p-cycle backup \cite{grover2003extending}, all node pair connections can be protected also.

Knowing the unicast or multicast requests a priori is often not possible.  This constraint makes protection against faults in those arbitrary communication paths a challenge.  An efficient all-nodes-to-all-nodes protection scheme supporting both unicast and multicast communication is necessary.

For efficiency and distributed control, it is common in distributed systems and algorithms to group nodes into intersecting sets referred to as quorum sets.  In this paper (and the prior, shorten conference version \cite{dlastine2014quorum}), it is shown that efficiency and distributed control can also be accomplished in optical network routing by applying the same established quorum set theory.  

The rest of the paper is organized as follows.  Sections \ref{sec:NetworkModel}, \ref{sec:light_trailDefinition}, and \ref{sec:cycleRouting} establish the network model, node communication, and path routing / fault tolerance.  In Section. \ref{sec:Quorums}, we apply the distributed efficiency of the quorum sets to routing optical cycles.  This method of route construction allows for the efficient handling of dynamic unicast and multicast communication requests. Sections \ref{sec:NSF_cycle_example} and \ref{sec:cycleLength} address routing and solution efficiencies.  Lastly, Section \ref{sec:FaultTolerance} analyzes the performance of our quorums set cycle routing techniques in the presence of network link faults.

\section{Network model} 
\label{sec:NetworkModel}
No two fiber-optic networks are the same.  Some stretch hundreds of kilometers, while other networks are contained within buildings or rooms.  Regardless of the physical environment, these optical circuits are depend\-ed upon for high-speed communications.  Thus, it is important to extract the network's critical components that affect its ability to deliver reliable, arbitrary point-to-point and multi-point communications.

These fiber-optic networks consist of several transmitters and receivers interconnected by fiber-optic cables.  As you might expect, transmitters and receivers are typically found together and generically called an optical node.  The cables form the links (i.e., edges) between those nodes, which leads to a convenient model of a network in terms of a graph $G = (V,E)$.  $V$ are the set of nodes in the network and $E$ are the set of edges. 

Edge $(a_i,a_j)$ is a fiber-optic link connecting nodes $a_i$ and $a_j$ in the network, where $a_i,a_j \in V$ and $(a_i,a_j) \in E$.  It is a general assumption that the same set of optical wavelengths are available on all edges in $E$.  The number of wavelengths available per optical fiber is dependent on the fiber-optic cables and the transmitter/receiver pairs.

\section{Light-trails} 
\label{sec:light_trailDefinition}
Lightpaths were a critical building block in the first optical communications, but required significant traffic engineering and aggregation to support point-to-point communication, or pay the penalty of low resource utilization on the fiber-optic link.  Lightpaths cannot support multicast traffic.  Light-trails were proposed in \cite{gumaste2003light,chlamtac2003light} as a solution to the challenges facing lightpaths and could be built using commercial off-the-shelf technology.  In the years since the introduction of light-trails, significant contributions have been made to enable adoption and advance the architecture \cite{dlastine2012fault,fang2004optimal,li2008multicast,zhang2011dynamic,dlastine2011ECBRA}.

Light-trails enable fast, dynamic creation of an unidirectional optical communication channel.  This communication channel, unlike prior lightpaths, allows for channel receive and transmit access to all connected nodes, making them more suitable for IP-centric traffic \cite{fang2004optimal}.  Point-to-point communications from an upstream node to a downstream node can be scheduled on the shared light-trail.  Similarly, an upstream node can multicast to any number of downstream nodes.  

A scheduling protocol is in place to avoid collisions within a light-trail and controls when nodes are able to transmit to downstream nodes.  The scheduling is generally assumed to occur over a control channel, which may or may not be separate from the shared optical fiber that is being used for the light-trail.

\begin{figure}[!tbp]
	\centering
	\includegraphics[width=2.5in]{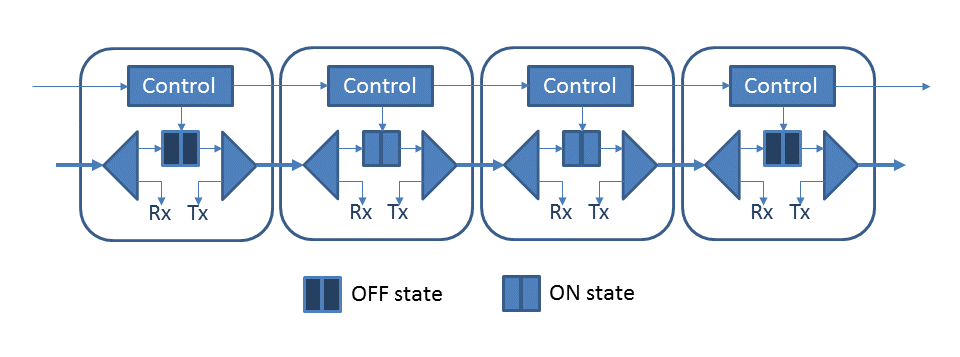}
	\caption{Four nodes in a light-trail architecture}
	\label{fig:lighttrail_arch}
\end{figure}

An example four-node light-trail can be seen in Figure \ref{fig:lighttrail_arch}.  Optical shutters allow for wavelength reuse within the network. Start and end nodes have their optical shutters in the \textit{off} state, while intermediate nodes have their optical shutters in the \textit{on} state.  This effectively isolates an optical signal to a specific light-trail and allows for reuse of optical wavelength(s) elsewhere in the network.

\begin{figure}[!tbp]
	\centering
	\includegraphics[width=2.5in]{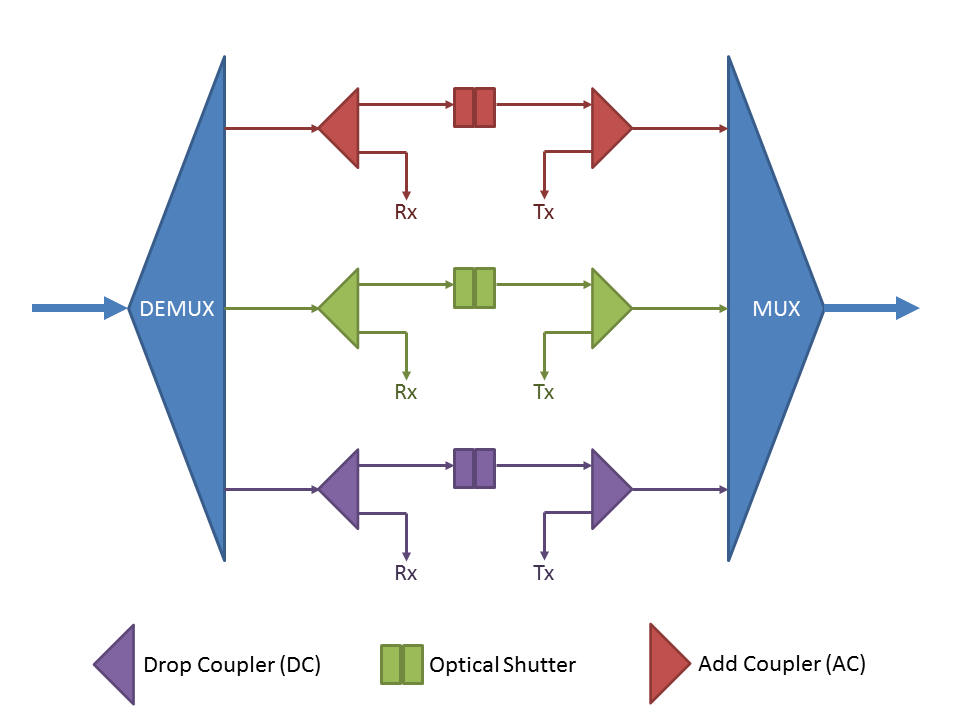}
	\caption{Example light-trail node structure}
	\label{fig:lighttrail_node}
\end{figure}

Nodes can receive from the incoming signal while the signal is simultaneously continuing to downstream nodes, sometimes referred to as a drop and continue function.  The node structure can be seen in Figure \ref{fig:lighttrail_node}.  DWDM fiber-optic networks use reconfigurable add/drop multiplexers (ROADMs) to demux incoming signals into separate wavelengths, then to mux the wavelengths before being output once again.  Each wavelength can separately support the light-trail architecture, allowing multiple light-trails to share the same edge in the network.  Next-generation ROADMs further increase add/drop and switching flexibility while reducing costs \cite{ji2010colorless}.  Early technology supported only a few wavelengths; however, the latest devices may support over 100 channels for over 1 Terabits/s \cite{agrawal2007nonlinear}. 

Light-trail communication is all optical and uses the same wavelength(s) from start to end node.  Being all optical avoids any energy inefficiencies and time delays associated with unnecessary Optical-to-Electrical-to-Optical (O/E/O) conversions at intermediate hops.  Transmissions within long haul networks, potentially passing through one or more nodes, used to be limited by the optical signal to noise ratio (OSNR).  In recent years, several advancements and mitigating techniques have allowed for this limitation to be reduced and in some cases completely removed.  One such advancement was the erbium- and ytterbium-doped optical fiber amplifier.  These amplifiers compensate for signal losses and allow for signals to travel thousands of kilometers \cite{agrawal2007nonlinear}.

\section{Light-trail, cycle routing, and fault-tolerance}
\label{sec:cycleRouting}
\begin{figure}[!tbp]
	\centering
	\includegraphics[width=2.5in]{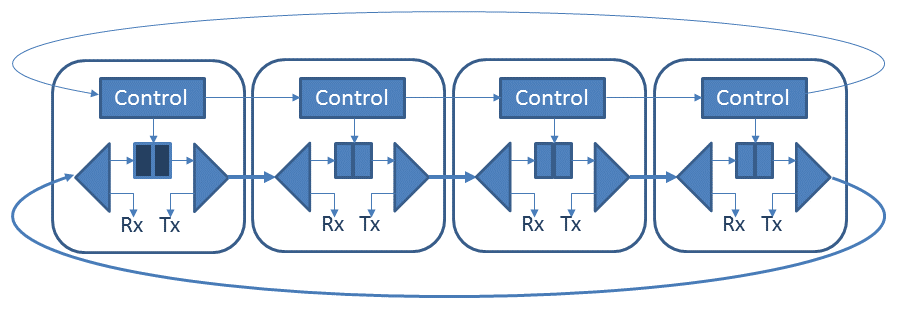}
	\caption{Cycle formed using the light-trail architecture}
	\label{fig:lighttrail_cycle}
\end{figure}

Point-to-point and multi-point traffic requests have a set of nodes $C = \{a_i,...,a_j\}$ that wish to communicate and need to be protected against network faults.  Establishing a primary and backup multicast path from every node to every other node in $C$ can be a waste of resources.  Several methods protect the path or links along the route through an independently found tree or cycle.  In this work, we utilize the light-trail architecture in the form of a cycle (Fig. \ref{fig:lighttrail_cycle}).  The bidirectional cycle will both route the multi-point request and protect it at the same time using fewer resources.

Figure \ref{fig:lighttrail_cycle} is simply a light-trail where the start and end node is the same node, referred to as the \textit{hub node}.  The hub node has its optical shutters in the \textit{off} state, while intermediate nodes have their optical shutters in the \textit{on} state.  The resources at each hub node can be utilized to allow all-to-all communication on the cycle using only one light-trail.  Traffic from a node to nodes downstream requires a single transmission.  Traffic from a node to an upstream node must undergo Optical-to-Electrical-to-Optical (O/E/O) conversion at the hub node and be transmitted on the light-trail a second time.

Alternately, we choose to set up two light-trails, one in each direction.  This enables upstream communications without the energy inefficiencies and time delays associated with O/E/O conversions.  It also has fault tolerance properties.

Additionally, if traffic in the network is expected to be significantly different than equal traffic between all node pairs, multiple light-trails can be used to support those specific cycles expecting heavier traffic.

Failures within an optical network are to be expected.  The generalization of p-cycle protection to allow for path and link protection was proposed by \cite{grover2003extending}.  P-cycle protection of unicast and multicast traffic networks requires preconfiguration, and the offline nature allows for the efficient cycles to be selected \cite{zhang2008performance,zhang2008optimizations}.  \cite{ramamurthy2003survivable} examined both path-based and link-based protection schemes in WDM networks.  The use of path-pair protection, link-based shared protection, spanning paths, and p-cycles to protect multicast sessions have all been proposed for WDM networks as well \cite{singhal2003provisioning,qing2005protecting,luo2006protecting,zhang2007applying,feng2008intelligent}. The Optimized Collapsed Rings (OCR) single link protection heuristic was developed to address the heterogeneous, part multicast / part unicast, nature of WDM traffic \cite{khalil2005pre}.

The multi-point cycle routing algorithm (MCRA) uses bidirectional cycles for fault tolerance and is capable of supporting SONET rings and p-cycles \cite{dlastine2012fault}.  Although finding the smallest cycle supporting the multi-point communication is NP-Complete, the authors were able to show that their heuristic performed within 1.2 times of the optimal cycle size.  ECBRA is a significant improvement of MCRA and outperforms the OCR heuristic \cite{dlastine2011ECBRA}.   

ECBRA heuristic balances optimality and speed, taking $O(\left|E\right|\left|C\right|^3)$ steps to find a close to optimal cycle.  First, a modified breadth first search is performed on each node in set $C$ of required communication nodes.  The goal is to find a shortest path in $G$ that also has the best ratio of nodes from set $C$ versus total nodes on the path.  The heuristic gives preference to paths with 2-degree $C$ nodes as these 2-degree nodes are required to be a part of the cycle.

To complete the cycle, a path from the sink node returning to the source node must be found.  No links may be used twice.  If all nodes in $C$ are in the cycle, then the cycle search is complete.  Otherwise, a third step is required to add any missing nodes.

If needed, the final step iteratively removes edges from the cycle and inserts paths through missing nodes in $C$.  Because insertion of the node can be cheaper by removing some links from the cycle rather than others, the optimal edge removal from the cycle and path replacement is computed for each missing node insertion.  

\section{Quorums} 
\label{sec:Quorums}
In distributed communication and algorithms, coordination, mutual exclusion, and consensus implementations have grouped $N$ nodes into small sets called quorums.  This organization of nodes can minimize communications in operations like negotiating access to a global resource. 

A quorums set minimally has the property that all quorums in the set must intersect.  Specifically for distributed implementations, it is also desirable that each node has equal work and equal responsibility within the quorums set \cite{maekawa1985algorithm}.  

Not every grouping of nodes into sets (quorums) will result in having these three properties, nor will the quorum sizes be minimal.  \cite{maekawa1985algorithm} proved the lower bound on the size of quorums set having these three properties.  Cyclic quorums sets with these properties, like other quorums sets in general, are difficult to find and require an exhaustive search \cite{maekawa1985algorithm,luk1997two}.  Quorums sets are proposed in this paper (and a conference paper \cite{dlastine2014quorum}) for cycle-based routing to efficiently support arbitrary point-to-point and multi-point communication.

\subsection{Defining quorums set}
\label{sec:Quorums:sub:Definition}
$A$ is a set of $N = \left|V\right|$ nodes.  A set $S_i$ is a subset of $A$.  When set $Q$ of subsets (Eq. \ref{eq:setQ}) covers all nodes in $A$ (Eq. \ref{eq:quorumUnion}) and all subsets also have non-empty intersections (Eq. \ref{eq:quorumIntersection}), then set $Q$ is called a quorums set.
\begin{align}
	A& = \{a_1,...,a_N\}\\
	S_i& = \{a_j, ...\},\quad j \in 1, 2, ..., N\\
	Q& = \{S_1,...,S_N\}\label{eq:setQ}\\
	\bigcup _{i=1}^NS_i& = \{a_1,...,a_N\} = A\label{eq:quorumUnion}\\
	S_i \cap S_j& \neq \emptyset, \quad \forall \;\;\; i,j \in 1,2,...,N\label{eq:quorumIntersection}
\end{align}

The lower bounds for the maximum individual quorum size (i.e., $\left|S_i\right|$) in a minimum set is $K$, where Equation \ref{eq:quorumSize} holds and $(K-1)$ is a power of a prime, proved through equivalence to finding a finite projective plane \cite{maekawa1985algorithm}. Additionally it is desirable that each quorum $S_i$ in the quorum set be of equal size (Eq. \ref{eq:equalWork}), such that there is equal work and it is desirable that each node be contained in the same number of quorums (Eq. \ref{eq:equalResponsibility}), such that there is equal responsibility.
\begin{align}
	N \leq K&(K-1) + 1\label{eq:quorumSize}\\
	\left|S_i\right| = K&, \quad \forall \;\; i \in 1, 2, ..., N\label{eq:equalWork}\\
	a_i \textrm{ is contained in } K& \, S_j\,'s,\quad \forall \;\;\; i \in 1, 2, ..., N\label{eq:equalResponsibility}
\end{align}

To the best of our knowledge, no efficient algorithm is known to find quorums of minimum size that maintain the properties used by \cite{maekawa1985algorithm}.  Cyclic quorums adhere to these properties and are based on cyclic block design and cyclic difference sets; however, searching for optimal cyclic quorum requires an exhaustive search \cite{luk1997two}.  Cyclic quorums are unique in that once the first quorum (Eq. \ref{eq:firstquorum}) is defined the remaining quorums can be generated via incrementing the entity ids (modulus to keep entity ids within bounds is not shown in Eq. \ref{eq:incrementquorum} for conciseness).  For simplicity assume $a_1 \in S_1$ without loss of generality (any one-to-one re-mapping of entity ids can result in this assumption).
\begin{align}
	S_1& = \{a_1, ..., a_j\}\label{eq:firstquorum}\\
	S_i& = \{a_{1+(i-1)}, ..., a_{j+(i-1)}\}\label{eq:incrementquorum}
\end{align}

For our work, we used the $N=4,...,111$ optimal cyclic quorums from a paper by Luk and Wong \cite{luk1997two}.

\subsection{Quorums sets for routing}
\label{sec:Quorums:sub:Routing}
It is important to establish quorums relationship to the network model $G = (V,E)$.  Quorums sets cover $N$ entities, in this case $N=\left|V\right|$ optical network nodes.  The number of entities, $N$, also defines the number of small subsets, i.e., quorums, that will be used in our solution.

In this paper (and a conference paper\cite{dlastine2014quorum}), the use of quorums for efficient point-to-point, multi-point, and all-to-all traffic requests in optical networks are proposed.  This is important because traffic in many optical networks is heterogeneous meaning the routing framework must be able to handle all types.

Point-to-point, multi-point, and all-to-all traffic can be routed through an optical network with $N$ cycles based on cyclic quorums.  The following breaks down the handling of different traffic request types on the quorum supporting rings.  Sections \ref{sec:Quorums:sub:Routing:sub:P2P} and \ref{sec:Quorums:sub:Routing:sub:MP2MP} expand upon a conference paper presented by Lastine and Somani \cite{dlastine2014quorum}.

\subsubsection{Point-to-point traffic}
\label{sec:Quorums:sub:Routing:sub:P2P}
Using cyclic quorums sets (see additional definitions in Sect. \ref{sec:Quorums:sub:Definition}) all possible node pairs occurred in at least one cycle.  This ensures for any dynamic point-to-point request there is at least one cycle that can support it.  

For many networks, node pairs will appear in multiple cycles.  When new requests arrive, they can select a route on the least loaded light-trail available.  Taking this one step further, if cycles with cyclical quorums are intentionally made larger than the minimum required, then it can be ensured that all networks have at least some opportunity for load balancing point-to-point connections.

\subsubsection{Multi-point traffic}
\label{sec:Quorums:sub:Routing:sub:MP2MP}
Optimally, if all multi-point participants belong to the same cycle, then one cycle can be used.  Realistically though, dynamic requests often will not be of this nature.  Requests will span multiple quorums and/or be larger than a single quorum cycle.  Hence in the worst case, no more than $K$ cycles are required to efficiently route and protect traffic for each multicast traffic request (more discussion on this bound in the broadcast traffic discussion, Sect. \ref{sec:Quorums:sub:Routing:sub:broadcast}).  

Additionally, in \cite{li2008multicast}  the problem of multicast in the general case for light-trails and for light-trail WDM networks is studied.  A way of transforming the general case to a minimum Steiner tree problem is given.  A polynomial time algorithm is given for the special case of light-trail WDM ring networks.  Work from this paper can be used to plan multicast requests that become known after the cycles based on quorums have been selected.

If we know of multicast requests the network will be serving a priori, we can number nodes to reduce the number of light-trails required to support the multicast.  The simplest case is when there is only one multicast that needs to be supported and the nodes involved in the multicast is equal or smaller than the number of nodes in a quorum.  In this case, we can renumber the nodes such that they are all in one cyclic quorum.

If there are two known multicasts, each not larger than a quorum, mapping may still be fairly simple.  If they both use some of the same nodes, there may be a pair of quorums that also has at least the same number of overlaps.  So two multicast could be accommodated by numbering nodes common to multicasts with numbers common to quorums, and then naming the rest of the nodes in the multicasts after nodes from one or the other quorum.

With each additional known multicast to support, the mapping to quorums can become more of a challenge.  We leave this as an open question on how to handle the general case of numbering nodes to handle multiple multicast requests known a priori.

\subsubsection{Broadcast traffic}
\label{sec:Quorums:sub:Routing:sub:broadcast}
Broadcast traffic is simply the worst case of a multicast traffic request.  The upper bound of requiring no more than $K$ cycles to route and protect up to broadcast traffic can described as follows.  In Section \ref{sec:Quorums:sub:Definition}, any entity will occur in at most $K$ quorums.  In those $K$ quorums, all other entities must be present in order to form necessary point-to-point pairs as described before (Sect. \ref{sec:Quorums:sub:Routing:sub:P2P}).  Hence, any optical node can communicate on all $K$ quorum cycles that it is a member and reach all other optical nodes, thereby efficiently serving any dynamic broadcast request.

\section{Efficiency analysis}
\label{sec:NSF_cycle_example}
The results presented in this section were adapted from results in a conference paper presented be Lastine and Somani \cite{dlastine2014quorum}.  

NSFNET is a 14 node 22 link network (Fig. \ref{fig:Networks:sub:nsfnet}).  At a size of 14 nodes, it has cyclic quorums of size five.  An improved version of ECBRA \cite{dlastine2014thesis} was used to find cycles that included the specified set of nodes in each quorum.

\paragraph{\textbf{Routing efficiency}}
Quorums for NSFNET are given in Table \ref{tbl:quorums_cycles_NSFNet_14}.  The first column numbers these quorums from 1 to 14.  The second column lists the actual quorums.  The third column lists cycles in the network that contain the quorums.  The fourth column provides the length (size) of each cycle as a convenience to the reader.  Cycle lengths fall in the range 7-11 links.  Their average length being 8.57 links and combined length is 120 links.

\begin{table}[!tb]
	\begin{center}
		\caption{ Cyclic quorums and cycles found in NSFNET }
		\label{tbl:quorums_cycles_NSFNet_14}
		\begin{tabular}{ |r|l|l|r| }
			\hline
			\# & 	Quorum 	           & Cycle                                 & Size \\ \hline
			1  &	1	2	3	4	8  & 1  4  12  13  8  3  2                 & 7    \\ \hline
			2  &	2	3	4	5	9  & 2  1  4  5  6  9  10  7  3            & 9    \\ \hline
			3  &	3	4	5	6	10 & 3  7  10  9  6  14  4  5              & 8    \\ \hline
			4  &	4	5	6	7	11 & 4  12  11  10  7  3  5  6  14         & 9    \\ \hline
			5  &	5	6	7	8	12 & 5  6  9  10  7  8  13  12  4          & 9    \\ \hline
			6  &	6	7	8	9	13 & 6  9  10  13  8  7  3  5              & 8    \\ \hline
			7  &	7	8	9	10	14 & 7  8  3  2  14  6  9  10              & 8    \\ \hline
			8  &	8	9	10	11	1  & 8  13  12  11  10  9  6  14  1  2  3  & 11   \\ \hline
			9  &	9	10	11	12	2  & 9  6  14  2  1  4  12  11  10         & 9    \\ \hline
			10 &	10	11	12	13	3  & 10  11  12  13  8  3  7               & 7    \\ \hline
			11 &	11	12	13	14	4  & 11  10  13  8  3  2  14  4  12        & 9    \\ \hline
			12 &	12	13	14	1	5  & 12  13  8  3  5  4  14  1  4          & 9    \\ \hline
			13 &	13	14	1	2	6  & 13  10  9  6  14  2  1  4  12         & 9    \\ \hline
			14 &	14	1	2	3	7  & 14  6  9  10  7  3  2  1              & 8    \\ \hline
			\multicolumn{3}{r|}{\textbf{Average:}}                             & \textbf{8.57} \\ \cline{4-4}
			\multicolumn{3}{r|}{\textbf{Total:}}                               & \textbf{120} \\ \cline{4-4}
		\end{tabular}
	\end{center}
\end{table}

Supporting all-to-all traffic using traditional point-to-point connections would be more expensive.  There are 91 possible node pairs.  With only 22 of the pairs connected via a direct link, it would take over 200 links to support all point-to-point communication paths.  The quorums set cycle routing required only 120 links.

\paragraph{\textbf{Resource efficiency}} 
Less hardware is required by the quorum solution as well.  The minimum number of transmitters and receivers at a node depends on the number of times it appears in the set of quorums.  From Section \ref{sec:Quorums:sub:Definition}, this lower bound is $K$ for optimal cyclic quorums.  In the example of the 14 node NSFNET, where $K = 5$, a node minimally needs only 5 transmitters and 5 receivers.  This is a significant saving over establishing lightpaths between all pairs of nodes which would require 13 transmitters and 13 receivers at each node.  

\section{Cycle length}
\label{sec:cycleLength}
The results presented in this section were adapted from results in a conference paper presented be Lastine and Somani \cite{dlastine2014quorum}.  

\subsection{Random graph solution cost analysis}
\label{sec:cycleLength:sub:RandomGraph}
Nodes are randomly placed inside a grid and links are randomly inserted with probability given by the Waxman model.   In general, the probability of a link existing depends on the distance between a node pair and the desired density of the network.  For context, graphs with a mean of approximately 2 edges per node are sparse in optical networking.  Whereas a mean of $log_2n$ edges per node would be dense in this context.

Graphs with various parameters had to be considered, in order to empirically approximate performance of our quorums set cycle routing on a general network.  Parameters considered were number of nodes (20, 35, 50, and 65), edge density (sparse, medium, and dense), and edge length (short and long).

Simulation results show that regardless of edge density or length, networks with more nodes intuitively required larger cycles in order to connect all nodes to each other.  Denser graphs allow for more route options for the cycle, which allows for smaller cycles to be formed including fewer non-quorum nodes in the cycle.  Lastly, graphs with higher probabilities for long edges on average had smaller cycles than graphs of the same size and density, but shorter edges.  The intuition for this comes from the need to connect all nodes and longer edges would allow ``skipping'' a few middle nodes when connecting to a node across the network.

A rough measurement of routing efficiency is produced by computing the ratio of average total nodes in a cycle to the required quorum nodes.  The quorums for graphs of size (20, 35, 50, and 65) nodes, respectively, contain (6, 7, 8, and 9) nodes.  As expected, simulation results show that the denser graphs had more efficient routes.  Also graphs with longer edges tended to have more efficient routes.  Interestingly though, increases in network size did not always result in decreases in efficiency as one might expect.  This could be partly due to the nature of which nodes are included in which quorum, which motivates the next set of results on whether altering the node number scheme in a graph could impact solution quality. 

\subsection{Impact of node numbering analysis}
Choosing of a node id for a particular node is likely a result of convenience rather than science.  Examining several random node numberings for graphs found in literature (Fig. \ref{fig:Networks}), the impact of node numbering on routing efficiency was investigated.

Total links used for routing (routing efficiency) did vary depending on the node numbering.  All networks experienced greater than 4\% difference between their observed minimum and maximum total links used in their randomly numbered graphs.  That is like having an extra cycle routed in each optical network, but not needing it.

The biggest result from this investigation was that all random node numberings for all networks had worse routing efficiency than the original node numberings in Figure \ref{fig:Networks}.  There is some intuitiveness to this result.  The cyclic quorums in \cite{luk1997two} tend to include several nodes with closely spaced id numbers and include fewer id numbers spaced further away.  When the original node numberings of the networks had nodes close to one another with closely spaced id numbers, this worked to the advantage of forming smaller cycles based on quorums.

\section{Fault tolerance analysis}
\label{sec:FaultTolerance}
\begin{figure}[!tbp]
	\centering
	\includegraphics[width=2.5in]{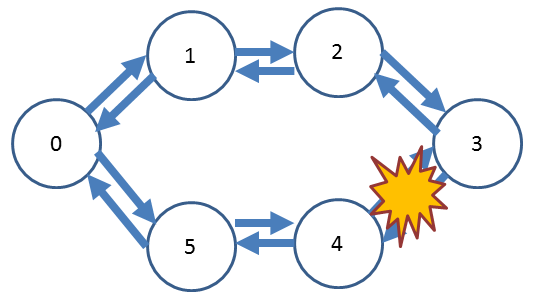}
	\caption{Example route fault tolerance using light-trails}
	\label{fig:lighttrailfail}
\end{figure}

Much of the discussion so far has addressed non-fault scenarios, specifically showing efficiencies in performing unicast, multicast, and broadcast operations using cycles based on quorums sets.  Optical networks are highly depended upon.  The fault tolerance aspect of this route design is important and a paired light-trail implementation is used to address this.  When a link fails, both cycles break.  In the Figure \ref{fig:lighttrailfail} example, hub node $3$ would no longer have a downstream edge to node $4$; however, upstream communication can be used to still reach node $4$.  

\subsection{Fault model}
\label{sec:FaultTolerance:sub:model}
The fault model assumed for our work is the single edge failure.  While a simple model, it does cover most real single fault scenarios.  

The most direct fault to consider is the optical link fault.  This occurs when a link is broken, like planned maintenance or the accidental severing during land excavation.  Modeling link faults as a single edge failure is straightforward.

Each modeled node needs a pair of transmitters and receivers for each occurrence in a cycle.  These pairs of devices can fail too.  Short of a natural disaster, pairs will likely fail independently of one another.  When a transmitter/receiver pair fails within a modeled node, the affect on the global network is similar to that link failing.  Modeling as a single edge failure, while not an exact fault mapping, is an appropriate abstraction.

We do not examine multiple simultaneous faults in our simulation models. 

%
\begin{figure*}[!tb]
	\centering
	\subfloat[NSFNET, 14-Node/22-Link][NSFNET\\14-Node/22-Link]{\includegraphics[width=2.5in]{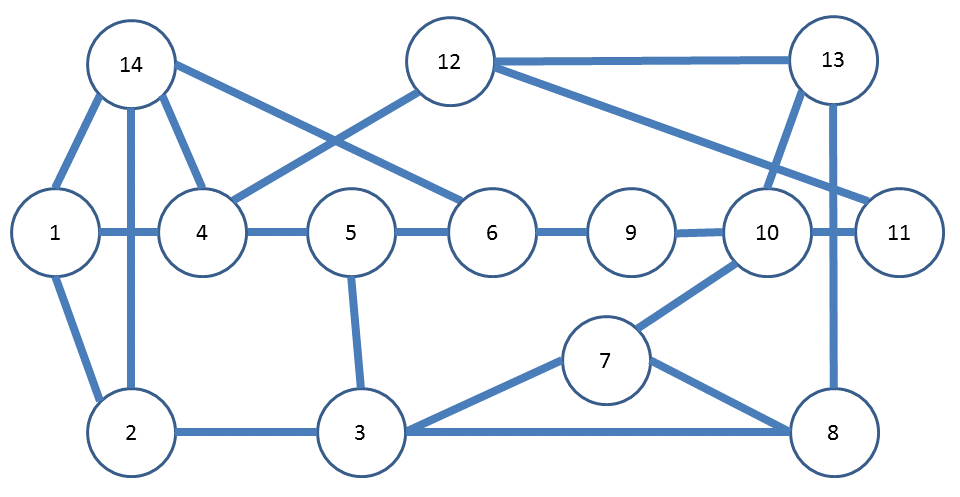}
		\label{fig:Networks:sub:nsfnet}}
	\hfil
	\subfloat[ARPANET, 20-Node/31-Link][ARPANET\\20-Node/31-Link]{\includegraphics[width=2.5in]{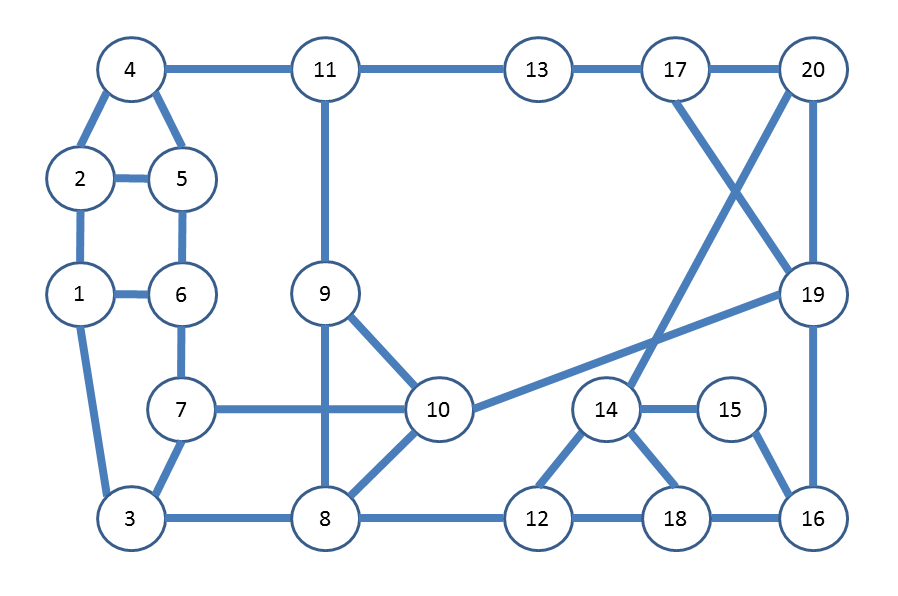}
		\label{fig:Networks:sub:arpanet}}
	\vskip\baselineskip
	\subfloat[American Backbone \cite{tang2011multicast}, 24-Node/43-Link][\centering American Backbone\cite{tang2011multicast}\par 24-Node/43-Link]{\includegraphics[width=2.5in]{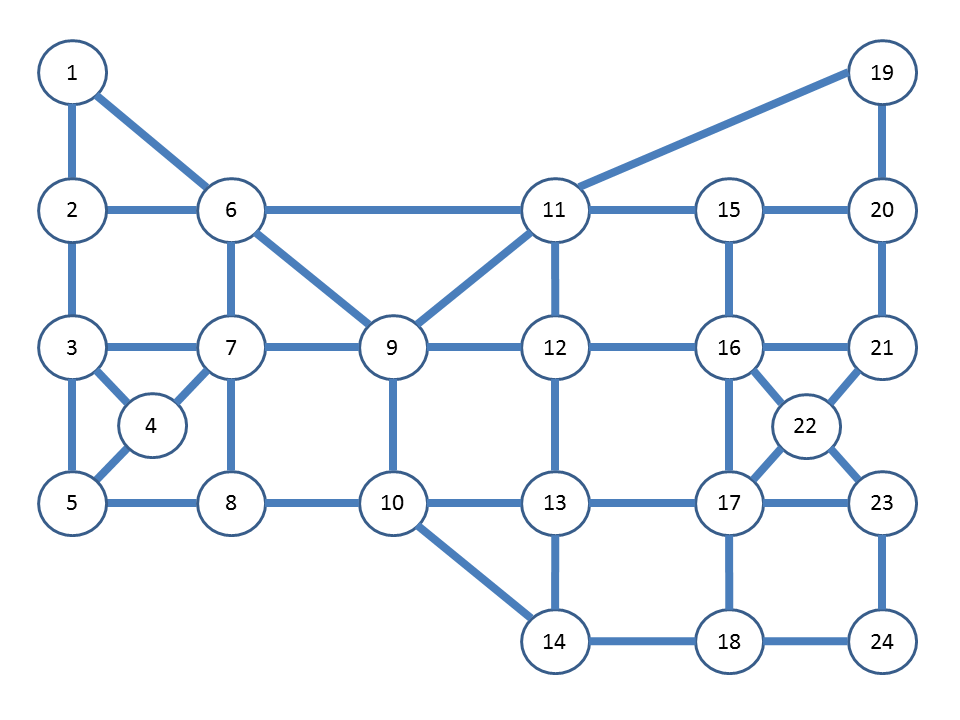}
		\label{fig:Networks:sub:american}}
	\hfil
	\subfloat[Chinese Backbone \cite{tang2011multicast}, 54-Node/103-Link][\centering Chinese Backbone\cite{tang2011multicast}\par 54-Node/103-Link]{\includegraphics[width=2.5in]{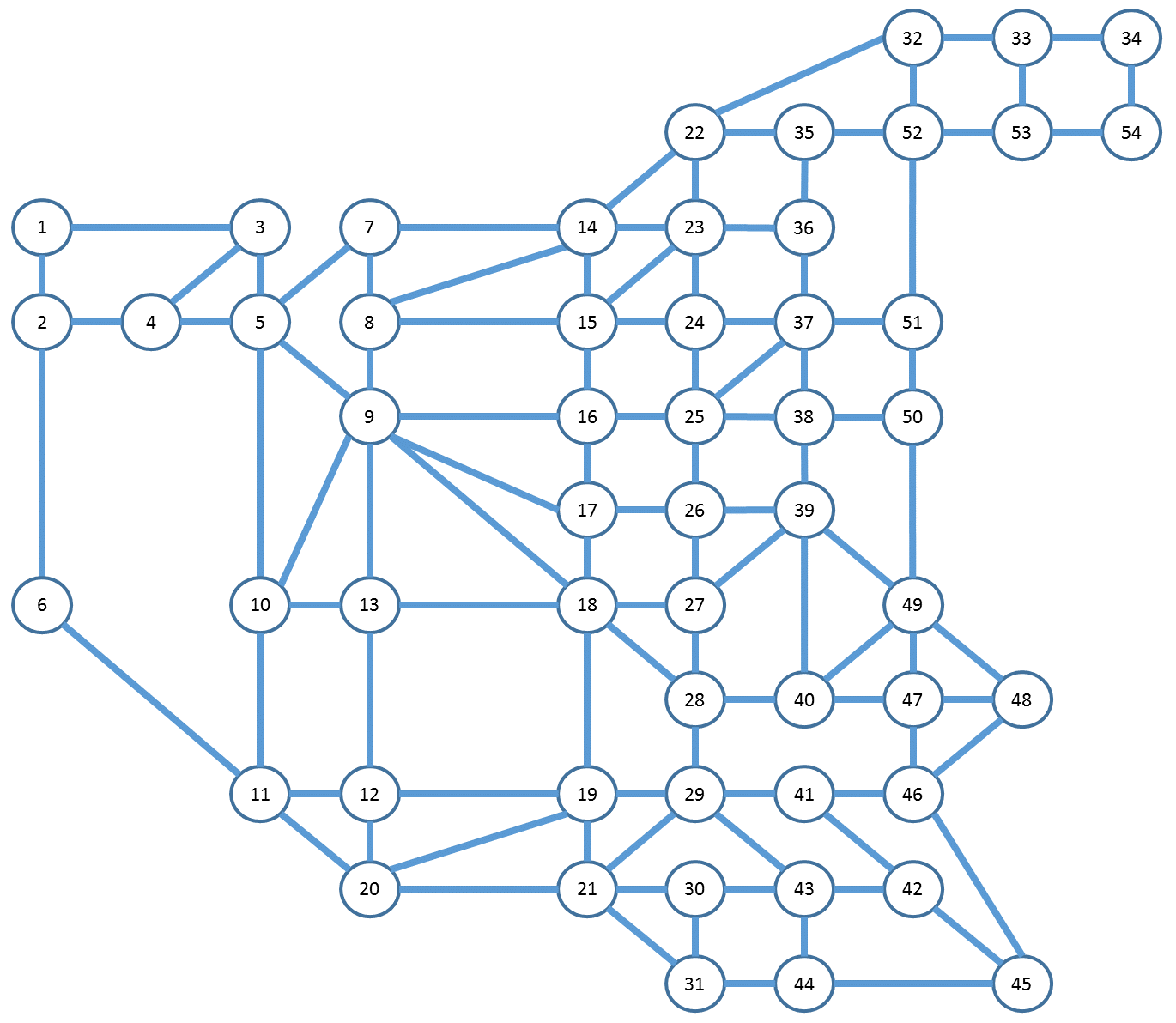}
		\label{fig:Networks:sub:chinese}}
	\caption{Networks used for simulations}
	\label{fig:Networks}
\end{figure*}

\subsection{Paired cycle fault simulation}
\label{sec:FaultTolerance:sub:pairedSim}
Maintaining the ability to serve all dynamic point-to-point traffic requests despite fault is important.  We examined the fault tolerance of the NSFNET, ARPANET, American backbone, Chinese backbone networks (Fig. \ref{fig:Networks}). For each of these four networks, we created 1000 random numbering schemes for the nodes and found cycles to support cyclical quorums given in \cite{luk1997two}.  

To model the fault, we simulate the failure of each edge, $e \in E$, in the network model, $G=(V,E)$.  We then examine the network's ability to serve all potential point-to-point requests by counting pairs of nodes that would be able to communicate and conversely those pairs that are unable to communicate.  In the non-fault case, all nodes can unidirectionally communicate with all other nodes for a total of $\left|V\right|*(\left|V\right|-1)$ pairs.

The results in Table \ref{tbl:PairedFaultSimulation} show acceptable fault tolerance performance.  The mean number of missing communication pairs is less than 3 (95\% CI).  This means that any edge fault in the network will typically be recoverable and few point-to-point connections will experience connection loss when using quorums set-based cycle routing.  

\begin{table*}[!tb]
	\begin{center}
		\caption{Paired quorum cycle fault simulation results}
		\label{tbl:PairedFaultSimulation}
		\begin{tabular}{|l|r|r|r|r|r|r|}
			\hline
			&                                     
			& 
			\multicolumn{1}{c|}{\textbf{Total}} & 
			\multicolumn{3}{c|}{\textbf{Missing Pairs}} & 
			\multicolumn{1}{c|}{\textbf{Fault}} \\ \cline{4-6}
			
			\textbf{Network} & 
			\multicolumn{1}{c|}{\textbf{Nodes}} & 
			\multicolumn{1}{c|}{\textbf{Pairs}} & 
			\multicolumn{1}{c|}{\textbf{High}} & 
			\multicolumn{1}{c|}{\textbf{Mean (95\% CI)}} & 
			\multicolumn{1}{c|}{\textbf{Low}} & 
			\multicolumn{1}{c|}{\textbf{Coverage (\%)}} \\ \hline
			NSFNET           & 14               &  182 & 12 & 0.93644 $\pm$ 0.02070 & 0 & 99.485 \\ \hline
			ARPANET          & 20               &  380 & 16 & 0.76051 $\pm$ 0.01715 & 0 & 99.800 \\ \hline
			American         & 24               &  552 & 26 & 2.05273 $\pm$ 0.02812 & 0 & 99.628 \\ \hline
			Chinese          & 54               & 2862 & 56 & 2.77809 $\pm$ 0.02400 & 0 & 99.903 \\ \hline
		\end{tabular}
	\end{center}
\end{table*}

To calculate fault coverage in this scenario, we calculated the mean connected pairs divided by the total pairs.
\[1-\frac{Mean\: Missing\: Pairs}{Total\: Pairs}\]
The results of our simulation showed that greater than 99\% fault coverage can typically be expected.

\subsection{Improving fault-tolerance}
\label{sec:FaultTolerance:sub:Improving}
Many of the simulated networks typically performed well, but ideally we would like to see the fewer missing pairs per fault case.  A reduction in the highest missing pairs observed is also important.

\subsubsection{Additional cycle fault protection}
Recall from Figure \ref{fig:lighttrail_cycle} in Section \ref{sec:cycleRouting} that hub nodes in cycles have their optical shutters in the \textit{off} state.  When a fault occurs like in Figure \ref{fig:lighttrailfail}, it is possible that the hub node is $0$ and the optical shutters could prevent a necessary communication path between $3$ and $4$.

\begin{figure}[!tbp]
	\centering
	\includegraphics[width=2.5in]{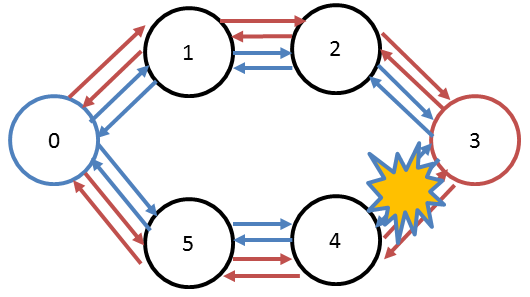}
	\caption{Quad light-trails to provide additional cycle fault protection}
	\label{fig:quadlighttrailfail}
\end{figure}

We experimented with adding an additional pair of cycles to form quad cycles.  The pair has its hub node directly across from the original pair's hub node, i.e., at position $\left\lfloor\frac{Cycle Length}{2}\right\rfloor$.  In the Figure \ref{fig:quadlighttrailfail} example, hub nodes $0$ (inner blue light-trail) and $3$ (outer red light-trail) are across from one another. Node $3$ still does not have a downstream edge to node $4$ on either the inner or outer cycle, but there does exist an upstream path on the outer cycle to node $4$.  

\begin{table*}[!tb]
	\begin{center}
		\caption{Quad quorum cycle fault simulation results}
		\label{tbl:QuadFaultSimulation}
		\begin{tabular}{|l|r|r|r|r|r|r|}
			\hline
			&                                     
			& 
			\multicolumn{1}{c|}{\textbf{Total}} & 
			\multicolumn{3}{c|}{\textbf{Missing Pairs}} & 
			\multicolumn{1}{c|}{\textbf{Fault}} \\ \cline{4-6}
			
			\textbf{Network} & 
			\multicolumn{1}{c|}{\textbf{Nodes}} & 
			\multicolumn{1}{c|}{\textbf{Pairs}} & 
			\multicolumn{1}{c|}{\textbf{High}} & 
			\multicolumn{1}{c|}{\textbf{Mean (95\% CI)}} & 
			\multicolumn{1}{c|}{\textbf{Low}} & 
			\multicolumn{1}{c|}{\textbf{Coverage (\%)}} \\ \hline			
			NSFNET           & 14               &  182 &  6 & 0.09130 $\pm$ 0.00574 & 0 & 99.950 \\ \hline
			ARPANET          & 20               &  380 &  6 & 0.08710 $\pm$ 0.00496 & 0 & 99.977 \\ \hline
			American         & 24               &  552 & 12 & 0.28661 $\pm$ 0.00833 & 0 & 99.948 \\ \hline
			Chinese          & 54               & 2862 & 20 & 0.52939 $\pm$ 0.00768 & 0 & 99.982 \\ \hline
		\end{tabular}
	\end{center}
\end{table*}

The results in Table \ref{tbl:QuadFaultSimulation} show that most of the networks had the mean number of missing pairs improved by an order of magnitude (95\% CI).  Any edge fault in the quad cycle network will typically be recoverable and fewer point-to-point connections will experience connection loss than with only the paired cycles.  This translates into higher fault coverage values as well.  Similarly the edge faults that generated the highest missing pairs observed in the simulated networks also decreased by $50\%$ or more.

Adding the additional pairs of cycles and the still having missing communication pairs may have been a bit surprising.  While the additional cycles significantly helped in fault reductions, the underlying missing pairs come from the intermediate nodes common to both cycles.  Depending on the failing edge, these nodes may not have a downstream or upstream path on any of the cycles to nodes on the opposite half of the cycle.

\subsubsection{Modifying the cycle routing algorithm}
Rather than addressing the missing point-to-point communication pairs with additional cycles, we could try to address the underlying cause of the missing pairs by changing the cycles themselves.  Here, we briefly outline additional algorithm steps that could be added to ECBRA. 

In Section \ref{sec:Quorums:sub:Routing:sub:P2P}, every point-to-point request was supported by at least one quorum cycle.  Being opportunistic, even if one cycle is experiencing a fault, there may be another quorum cycle that also is supporting that point-to-point request.  Hence, quantifying missing pairs and attempting to compensate should only occur once all cycles for a graph's quorums set is found.  

To enumerate the missing pairs, simulate the failure of each edge and examine the network's ability to serve all potential point-to-point requests by counting pairs of nodes that would be able to communicate and conversely those pairs that are unable to communicate.  Given results in Section \ref{sec:FaultTolerance:sub:pairedSim} the mean number of missing pairs is expected to be less than 3 per fault edge (95\% CI). Form a \textit{Missing-Pair} tuple $t = <\{s,d\},e>$, where $s$ is the source and $d$ is the destination of the missing communications pair and $e$ is the edge whose failure was being simulated.

Every pair missing can be isolated back to a cycle responsible for the pair (Sect. \ref{sec:Quorums:sub:Routing:sub:P2P}).  For each tuple $t$, remove the faulty edge $e$ in the responsible cycle for pair $\{s,d\}$.  To complete the cycle, find the shortest path between the two disconnected nodes, such that cycle edges are not reused and avoiding the use of $e$.  Once all Missing-Pair tuples have been processed, repeat the enumeration and removal of missing pairs once again.  Repetition is required to confirm that in the process of addressing one Missing-Pair tuple that another new missing pair was not added.

There is a possibility that a cycle may have to use edge $e$.  An example would be a node with only two edges and the node being a member of the quorum for that cycle.  This is a known limitation and is a challenge in general for establishing appropriate fault tolerance for communications to/from this node, as well as, through this node serving others.  This limitation is discussed further in \cite{dlastine2011ECBRA}.

\section{Conclusion}
\label{sec:Conclusion}
Many optical networks face heterogeneous communication requests requiring topologies to be efficient and fault tolerant.  We show efficiency and distributed control can also be accomplished in optical network routing by applying the quorums set theory.

Specifically we show cyclic quorums set routing have all pairs of nodes occurring in one or more quorums.  This allows for efficient, arbitrary unicast and multicast communications requests that do not need to be known or even modeled a priori. Even efficient broadcasts to all network nodes are possible by a node broadcasting to all quorum cycles to which it belongs.

We analyzed the fault tolerance of our quorums set routing approach.  In the presence of network link faults, greater than 99\% average coverage enables the continued operation of nearly all arbitrary unicast and multicast requests in the network.  Adding additional cycles showed an order of magnitude reduction in mean missing point-to-point pairs for most networks (95\% CI).  Finally, to address faults at their source, an augmentation to the ECBRA cycle finding algorithm was proposed.

\section{Future work}
\label{sec:FutureWork}
In this paper, ECBRA was used to route each of the quorums-based cycles.  It was shown that the quorums set approach required far fewer links to accomplish the routing of all-to-all traffic, when compared to using point-to-point connections.  

As an unintended benefit, some quorums sets result in node pairs occurring in more than one quorums-based cycle.  It was these occurrences of node pairs multiple times that motivates us to examine whether redundant node pairs could be generated intentionally.  Predictably redundant pairs can improve the dependability of the optical network, by guaranteeing even if a cycle failed, all node pairs for point-to-point and multi-point communications could still be present in the network.


%
%

\begin{acknowledgements}
The authors would like to acknowledge and thank Dr. David Lastine for his thoughtful discussion and his contributions and results presented in a conference paper \cite{dlastine2014quorum} that are adapted in our presentation in Sections \ref{sec:Quorums:sub:Routing:sub:P2P}, \ref{sec:Quorums:sub:Routing:sub:MP2MP}, \ref{sec:NSF_cycle_example}, and \ref{sec:cycleLength}.
\end{acknowledgements}

\bibliographystyle{IEEEtran} 
\bibliography{ICOCN_Extended_bib}

\begin{thebibliography}{10}
\providecommand{\url}[1]{#1}
\csname url@samestyle\endcsname
\providecommand{\newblock}{\relax}
\providecommand{\bibinfo}[2]{#2}
\providecommand{\BIBentrySTDinterwordspacing}{\spaceskip=0pt\relax}
\providecommand{\BIBentryALTinterwordstretchfactor}{4}
\providecommand{\BIBentryALTinterwordspacing}{\spaceskip=\fontdimen2\font plus
\BIBentryALTinterwordstretchfactor\fontdimen3\font minus
  \fontdimen4\font\relax}
\providecommand{\BIBforeignlanguage}[2]{{%
\expandafter\ifx\csname l@#1\endcsname\relax
\typeout{** WARNING: IEEEtran.bst: No hyphenation pattern has been}%
\typeout{** loaded for the language `#1'. Using the pattern for}%
\typeout{** the default language instead.}%
\else
\language=\csname l@#1\endcsname
\fi
#2}}
\providecommand{\BIBdecl}{\relax}
\BIBdecl

\bibitem{dlastine2012fault}
D.~Lastine, S.~Sankaran, and A.~K. Somani, ``A fault-tolerant multipoint cycle
  routing algorithm (mcra),'' in \emph{Broadband Communications, Networks, and
  Systems}.\hskip 1em plus 0.5em minus 0.4em\relax Springer, 2012, pp.
  341--360.

\bibitem{grover2003extending}
W.~D. Grover and G.~Shen, ``Extending the p-cycle concept to path-segment
  protection,'' in \emph{Communications, 2003. ICC'03. IEEE International
  Conference on}, vol.~2.\hskip 1em plus 0.5em minus 0.4em\relax IEEE, 2003,
  pp. 1314--1319.

\bibitem{dlastine2014quorum}
A.~K. Somani and D.~Lastine, ``Optical paths supporting quorums for efficient
  communication,'' in \emph{Optical Communications and Networks (ICOCN), 2014
  13th International Conference on}.\hskip 1em plus 0.5em minus 0.4em\relax
  IEEE, 2014.

\bibitem{gumaste2003light}
A.~Gumaste and I.~Chlamtac, ``Light-trails: a novel conceptual framework for
  conducting optical communications,'' in \emph{High Performance Switching and
  Routing, 2003, HPSR. Workshop on}.\hskip 1em plus 0.5em minus 0.4em\relax
  IEEE, 2003, pp. 251--256.

\bibitem{chlamtac2003light}
I.~Chlamtac and A.~Gumaste, ``Light-trails: A solution to ip centric
  communication in the optical domain,'' in \emph{Quality of Service in
  Multiservice IP Networks}.\hskip 1em plus 0.5em minus 0.4em\relax Springer,
  2003, pp. 634--644.

\bibitem{fang2004optimal}
J.~Fang, W.~He, and A.~K. Somani, ``Optimal light trail design in wdm optical
  networks,'' in \emph{Communications, 2004 IEEE International Conference on},
  vol.~3.\hskip 1em plus 0.5em minus 0.4em\relax IEEE, 2004, pp. 1699--1703.

\bibitem{li2008multicast}
Y.~Li, J.~Wang, A.~Gumaste, Y.~Xu, and Y.~Xu, ``Multicast routing in
  light-trail wdm networks,'' in \emph{Global Telecommunications Conference,
  2008. IEEE GLOBECOM 2008. IEEE}.\hskip 1em plus 0.5em minus 0.4em\relax IEEE,
  2008, pp. 1--5.

\bibitem{zhang2011dynamic}
W.~Zhang, F.~Kandah, C.~Wang, and H.~Li, ``Dynamic light trail routing in wdm
  optical networks,'' \emph{Photonic Network Communications}, vol.~21, no.~1,
  pp. 78--89, 2011.

\bibitem{dlastine2011ECBRA}
A.~Somani, D.~Lastine, and S.~Sankaran, ``Finding complex cycles through a set
  of nodes,'' in \emph{Global Telecommunications Conference (GLOBECOM 2011),
  2011 IEEE}, Dec 2011, pp. 1--5.

\bibitem{ji2010colorless}
P.~N. Ji and Y.~Aono, ``Colorless and directionless multi-degree reconfigurable
  optical add/drop multiplexers,'' in \emph{Wireless and Optical Communications
  Conference (WOCC), 2010 19th Annual}.\hskip 1em plus 0.5em minus 0.4em\relax
  IEEE, 2010, pp. 1--5.

\bibitem{agrawal2007nonlinear}
G.~P. Agrawal, \emph{Nonlinear fiber optics}.\hskip 1em plus 0.5em minus
  0.4em\relax Academic press, 2007.

\bibitem{zhang2008performance}
F.~Zhang and W.-D. Zhong, ``Performance evaluation of p-cycle based protection
  methods for provisioning of dynamic multicast sessions in mesh wdm
  networks,'' \emph{Photonic Network Communications}, vol.~16, no.~2, pp.
  127--138, 2008.

\bibitem{zhang2008optimizations}
F.~Zhang, W.-D. Zhong, and Y.~Jin, ``Optimizations of p-cycle-based protection
  of optical multicast sessions,'' \emph{Lightwave Technology, Journal of},
  vol.~26, no.~19, pp. 3298--3306, 2008.

\bibitem{ramamurthy2003survivable}
S.~Ramamurthy, L.~Sahasrabuddhe, and B.~Mukherjee, ``Survivable wdm mesh
  networks,'' \emph{Journal of Lightwave Technology}, vol.~21, no.~4, p. 870,
  2003.

\bibitem{singhal2003provisioning}
N.~K. Singhal, L.~H. Sahasrabuddhe, and B.~Mukherjee, ``Provisioning of
  survivable multicast sessions against single link failures in optical wdm
  mesh networks,'' \emph{Journal of lightwave technology}, vol.~21, no.~11, p.
  2587, 2003.

\bibitem{qing2005protecting}
Y.~Qing and G.~Ning, ``Protecting dynamic multicast sessions in optical wdm
  mesh networks,'' in \emph{Information, Communications and Signal Processing,
  2005 Fifth International Conference on}.\hskip 1em plus 0.5em minus
  0.4em\relax IEEE, 2005, pp. 1187--1190.

\bibitem{luo2006protecting}
H.~Luo, H.~Yu, L.~Li, and S.~Wang, ``On protecting dynamic multicast sessions
  in survivable mesh wdm networks,'' in \emph{Communications, 2006. ICC'06.
  IEEE International Conference on}, vol.~2.\hskip 1em plus 0.5em minus
  0.4em\relax IEEE, 2006, pp. 835--840.

\bibitem{zhang2007applying}
F.~Zhang and W.-D. Zhong, ``Applying p-cycles in dynamic provisioning of
  survivable multicast sessions in optical wdm networks,'' in \emph{Optical
  Fiber Communication Conference}.\hskip 1em plus 0.5em minus 0.4em\relax
  Optical Society of America, 2007, p. JWA74.

\bibitem{feng2008intelligent}
T.~Feng, L.~Ruan, and W.~Zhang, ``Intelligent p-cycle protection for multicast
  sessions in wdm networks,'' in \emph{Communications, 2008. ICC'08. IEEE
  International Conference on}.\hskip 1em plus 0.5em minus 0.4em\relax IEEE,
  2008, pp. 5165--5169.

\bibitem{khalil2005pre}
A.~Khalil, A.~Hadjiantonis, G.~Ellinas, and M.~Ali, ``Pre-planned multicast
  protection approaches in wdm mesh networks,'' in \emph{Optical Communication,
  2005. ECOC 2005. 31st European Conference on}.\hskip 1em plus 0.5em minus
  0.4em\relax IET, 2005, pp. 25--26.

\bibitem{maekawa1985algorithm}
M.~Maekawa, ``An algorithm for mutual exclusion in decentralized systems,''
  \emph{ACM Transactions on Computer Systems (TOCS)}, vol.~3, no.~2, pp.
  145--159, 1985.

\bibitem{luk1997two}
W.-S. Luk and T.-T. Wong, ``Two new quorum based algorithms for distributed
  mutual exclusion,'' in \emph{Distributed Computing Systems, 1997.,
  Proceedings of the 17th International Conference on}.\hskip 1em plus 0.5em
  minus 0.4em\relax IEEE, 1997, pp. 100--106.

\bibitem{dlastine2014thesis}
D.~Lastine, ``Efficient communication using multiple cycles and multiple
  channels,'' Ph.D. dissertation, Iowa State University, 2014.

\bibitem{tang2011multicast}
L.~Tang, W.~Huang, M.~Razo, A.~Sivasankaran, M.~Tacca, and A.~Fumagalli,
  ``Multicast tree computation in networks with multicast incapable nodes,'' in
  \emph{High Performance Switching and Routing (HPSR), 2011 IEEE 12th
  International Conference on}.\hskip 1em plus 0.5em minus 0.4em\relax IEEE,
  2011, pp. 95--100.

\end{thebibliography}

%
%

\begin{authorbiography}{Cory_Kleinheksel_4x3ratio}{Cory J. Kleinheksel} \hspace{4pt}received the B.S. degree in Computer Engineering from Iowa State University in 2008.  He is a Ph.D. candidate at Iowa State University pursuing a degree in Computer Engineering.  He has received fellowships from the NSF Graduate Research Fellowship Program, IBM Ph.D. Fellowship Program, and Symbi GK-12 Fellowship Program at Iowa State University.  His research interests are in the area of big data communication and processing with emphasis on parallel and distributed systems.
\end{authorbiography}

\begin{authorbiography}{Arun-small-size-Feb2015}{Arun K. Somani} \hspace{4pt}is serving as Anson Marston Distinguished Professor and Philip and Virginia Sproul Professor of Electrical and Computer Engineering at Iowa State University. He also served as Scientific Officer for Govt. of India, New Delhi from 1974 to 1982 and as a faculty member at the University of Washington, Seattle, WA from 1985 to 1997. He earned his MSEE and PhD degrees in electrical engineering from the McGill University, Montreal, Canada, in 1983 and 1985, respectively.
	
Professor Somani’s research interests are in the area of computer system design and architecture, fault tolerant computing, computer interconnection networks, WDM-based optical networking, and reconfigurable and parallel computer systems. He has also served as IEEE distinguished visitor, IEEE distinguished tutorial speaker, and IEEE Communication Society distinguished visitor and has delivered several key note speeches, tutorials and distinguished and invited talks all over the world. He was elected a Fellow of IEEE for his contributions to theory and applications of computer networks in 1999 and as a Distinguished Engineer of ACM in 2006. He also has been elected a fellow of AAAS in 2012.
\end{authorbiography}

\end{document}